# Terahertz superconducting plasmonic hole array

**Zhen Tian,[1,2] Ranjan Singh,[1] Jiaguang Han,[2,3] Jianqiang Gu,[1,2]**
**Qirong Xing,[2] and Weili Zhang[1]***

[1] *School of Electrical and Computer Engineering, Oklahoma State University, Stillwater, Oklahoma 74078, USA*

[2] *Center for Terahertz waves and College of Precision Instrument and Optoelectronics Engineering, Tianjin University, P. R. China,*

[3] *Department of Physics, National University of Singapore, 2 Science Drive 3, Singapore 117542, Singapore*

**Corresponding author: weili.zhang@okstate.edu*

**Abstract**

We demonstrate a superconductor array of subwavelength holes with active thermal control over the resonant transmission induced by surface plasmon polaritons. The array was lithographically fabricated on high temperature YBCO superconductor and characterized by terahertz-time domain spectroscopy. We observe a clear transition from a virtual excitation of the surface plasmon mode to a real surface plasmon mode. The highly controllable superconducting plasmonic crystals may find promising applications in the design of low-loss, large dynamic range amplitude modulation, and surface plasmon based terahertz devices.

The discovery of extraordinary transmission of light through gratings of subwavelength holes has ignited great interest in the field of surface plasmon optics (plasmonics) over the past few years from both the basic science and an applied physics point of view [1-3]. Plasmonics is predicted to have a strong future in high technology world for the design and development of next generation electronic and photonic chips [4]. To realize this promise, the bigger challenge is to actively control the surface plasmon polaritons (SPPs). It has been demonstrated that the SPP-assisted light propagation can be controlled through thermal, optical, and electrochemical means in the visible and near-infrared frequency regimes [5-7]. At terahertz frequencies, semiconductors are efficient materials for active plasmonics. Thermal, electrical, magnetic and optical switching of terahertz SPPs has been extensively demonstrated in recent years, allowing active control of the surface plasmon resonance [8-14]. The semiconductor and the metal used in fabricating plasmonic devices suffer from the limitation of higher losses. This drawback can be lifted by the use of high transition temperature (high-Tc) superconductor (HTS) thin films. The cuprate superconductors are highly anisotropic material and have multilayer superconducting $Cu_2O$ planes with interlayer tunneling of Cooper pairs between them, which introduces the *c*-axis Josephson plasma resonance (JPR). The JPR usually lies in the microwave and terahertz spectral regions [15]. Recent experimental and theoretical works have predicted the existence of surface waves in the HTS films [16-19].

In this Letter, we demonstrate the existence and active thermal control of SPPs in periodic subwavelength hole array made up of high-Tc YBCO by using terahertz-time domain spectroscopy (THz-TDS) [20]. In the hole array, due to *c*-axis Josephson plasma frequency, we observed a sharp transition between a virtual excitation type SPP mode and a real SPP mode accompanied with enhanced transmission amplitude modulation.

The sample is made from a commercial (THEVA, Germany) 280 nm-thick YBCO film which typically has an 86 K transition temperature and 2.3 MA/cm$^2$ critical current density grown on a 500 μm thick sapphire substrate. By using conventional photolithographic [21, 22] exposure we patterned a 3-μm-thick negative photo-resist, NR7-3000P film into hole shape on the YBCO film as the protective layer. The sample wan then wet etched in 0.04% nitric acid to remove YBCO from other parts of the wafer that did not have the photoresist protection followed by a lift-off process in acetone. The optical images of the hole array is illustrated in Fig. 1. The dimension of a unit hole is 65 × 50 μm$^2$ with a lattice constant of 100 μm.

The measurements were performed at room temperature and then gradually cooled down to 86 K in a cryo-assisted THz-TDS system used in previous measurement [23]. Figure 2 illustrates the measured normalized transmission response of YBCO hole array at normal incidence at 297, 183, 133 and 86 K, respectively. We observed two resonance modes at 0.85 and 1.16 THz with peak amplitudes of 0.69 and 0.51, respectively, which correspond to the $[\pm1,0]$ and $[\pm1,\pm1]$ modes of the surface wave. As the hole array is cooled down, both resonances gradually blue-shift with an increase in their respective peak amplitude. Finally, at the critical temperature of 86 K, the $[\pm1,0]$ resonance mode shifts to 0.88 THz with a peak amplitude 0.75.

The anisotropic dielectric function of YBCO can be written as [15]

$$\varepsilon_c(\omega) = \varepsilon_\infty^c (1 - \frac{\omega_{pc}^2}{\omega^2} + \frac{4\pi i \sigma_c}{\varepsilon_\infty^c \omega}),$$
$$\varepsilon_{ab}(\omega) = \varepsilon_\infty^{ab} (1 - \frac{\omega_{pab}^2}{\omega^2}), \tag{1}$$

where $\omega_{pc}$ and $\omega_{pab}$ are the out-of-plane and in-plane plasma frequency, respectively. The plasma frequency along *c* axis is also called Josephson plasma frequency, which usually lies in the microwave and terahertz regions. Since the critical current density $J_c$ is known, the Josephson plasma frequency $\omega_{Jp}$ can be evaluated using the following equation [24],

$$\omega_{Jp} = \sqrt{\frac{2eJ_c d}{\hbar \varepsilon_0 \varepsilon_c}}, \tag{2}$$

where $J_c$ = 2.3 MA/cm$^2$ is the critical current density, *d* is the interlayer spacing (≈ 1 nm), $\varepsilon_c$ = 15 is the dielectric constant of YBCO. Based on Eqs. (1) and (2), the estimated Josephson plasma frequency is 3.4 THz at the critical transition temperature and this value is consistent with that in the previous literature [25]. For temperatures higher than the critical transition value, the superconducting carriers will disappear and the YBCO film along *c* axis can be treated as a dielectric material. The plasma frequency along the *a-b* plane can be measured directly from the normal incidence transmission and is found to have a value close to that in the near-infrared regime. It should be noted that the structure of YBCO allows high conduction in Copper planes, thus confining the conductivity to the *a-b* planes giving rise to large anisotropy in the transport properties. The normal conductivity in the *c*- axis is an order of magnitude lower than that in the *a-b* plane.

Since YBCO is a highly anisotropic material, the dispersion relation of surface polaritons can be expressed as [26]

$$k^2 = \frac{\omega^2}{c^2}\varepsilon_s\varepsilon_c\frac{\varepsilon_{ab}-\varepsilon_s}{\varepsilon_{ab}\varepsilon_c-\varepsilon_s^2}, \tag{3}$$

where $\varepsilon_s$ is the dielectric constant of the surrounding substrate, $\varepsilon_{ab}$ and $\varepsilon_c$ are the dielectric function of YBCO along the *a-b* plane and *c* axis, respectively. This dispersion relation is consistent with that in the isotropic material if $\varepsilon_{ab}=\varepsilon_c$. For YBCO, however, $\varepsilon_{ab}\neq\varepsilon_c$, thus two cases arise. When the temperature is lower than the critical temperature of 86 K, $\varepsilon_{ab}<0$, $\varepsilon_c<0$ and $|\varepsilon_{ab}|\gg|\varepsilon_c|$, the dispersion relation can be simplified as $k^2=(\omega^2/c^2)\times\varepsilon_s$. In this case, the terahertz wave can be coupled to the real SPP mode which is similar to the case in metallic hole array. As the temperature soars above the critical transition, $\varepsilon_{ab}<0$, $\varepsilon_c>0$ and $|\varepsilon_{ab}|\gg\varepsilon_c$, the dispersion relation again simplifies to the same expression, $k^2=(\omega^2/c^2)\times\varepsilon_s$, but the surface mode excited here is a virtual SPP mode which occurs only for small magnitude of wave vector. Furthermore, the virtual excitation type SPPs must always be driven by the associated electromagnetic wave. When the associated electromagnetic wave is removed, the virtual SPP mode will disappear immediately [27]. Figure 2 clearly reveals that all the measured transmission resonances are caused by the virtual excitation of SPPs above 86 K.

The critical behaviors of HTS are very susceptible to defects and small changes in the chemical composition and structure which present a challenge in superconductor materials comprising of four or more elements. The YBCO film on sapphire substrate does not seem to reach the transition temperature at 86 K in our measurements. This may be due to material imperfections such as cracks, voids, and dislocations. The substrate crystalline structure must be ideally matched to the superconductor if the thin layer of superconductor is to perform predictably by minimizing the flaws and dislocations. The critical current in the superconductor is also strongly limited by the high angle grain boundaries.

We extracted the real conductivity of the YBCO film at 297, 183, 133, and 86 K by measuring the direct transmission and phase shift of terahertz waves through the film. In order to verify the experimental results, we simulated the terahertz transmission through the YBCO hole array using the measured real conductivity up to 86 K and the complex conductivity value at 51.5 K taken from literature [28,29]. The simulation tool used is CST microwave studio and the results are shown in Fig. 3(a). Due to increase in the ratio of the real to the imaginary dielectric constant [30], the peak amplitude transmission caused by the virtual excitation type SPP becomes larger gradually and at superconducting state the transmission peak caused by a real SPP nearly reaches 1 demonstrating extremely low loss. The frequency blue-shift is caused by the increase in real part of dielectric function [12]. The dramatic change in amplitude transmission can be seen in Fig. 3(b) where the transmission at resonance suddenly increases by almost 30% due to the formation of superconducting cooper pairs below the transition temperature. Before the superconductor reaches its transition temperature, the change in transmission from room temperature up to the critical temperature is only ~5%.

In summary, an active control over terahertz resonant transmission with amplitude modulation of 35% and sharp resonance strengthening has been achieved in periodic subwavelength YBCO hole array by cooling it down to temperatures below the superconducting transition temperature. Such plasmonic superconducting structures would open up promising avenues for the design and development of low-loss, large dynamic range amplitude modulator and temperature-controlled terahertz devices.

The authors thank J. Wu and J. Zhang for their support and discussions. This work was partly supported by the U.S. National Science Foundation, the China Scholarship Council, the National Key Basic Research Special Foundation of China (Grant Nos. 2007CB310403 and 2007CB310408), the Tianjin Sci-Tech Support Program (Grant No. 8ZCKFZC28000), the MOE Academic Research Fund of Singapore, and the Lee Kuan Yew Fund.

## Figure Captions

**FIG. 1** (color online). Optical images of (a) a periodic subwavelength YBCO hole array on sapphire substrate, and (b) a unit cell with geometric dimensions.

**FIG. 2** (color online). Measured amplitude transmission response of the YBCO hole array at 297, 183, 133 and 86K, respectively, at normal incidence.

**FIG. 3** (color online). Simulations for (a) terahertz amplitude transmission of the subwavelength YBCO hole array, and (b) amplitude transmission peak percentage at varying temperatures. The dotted lines are to guide the eye.

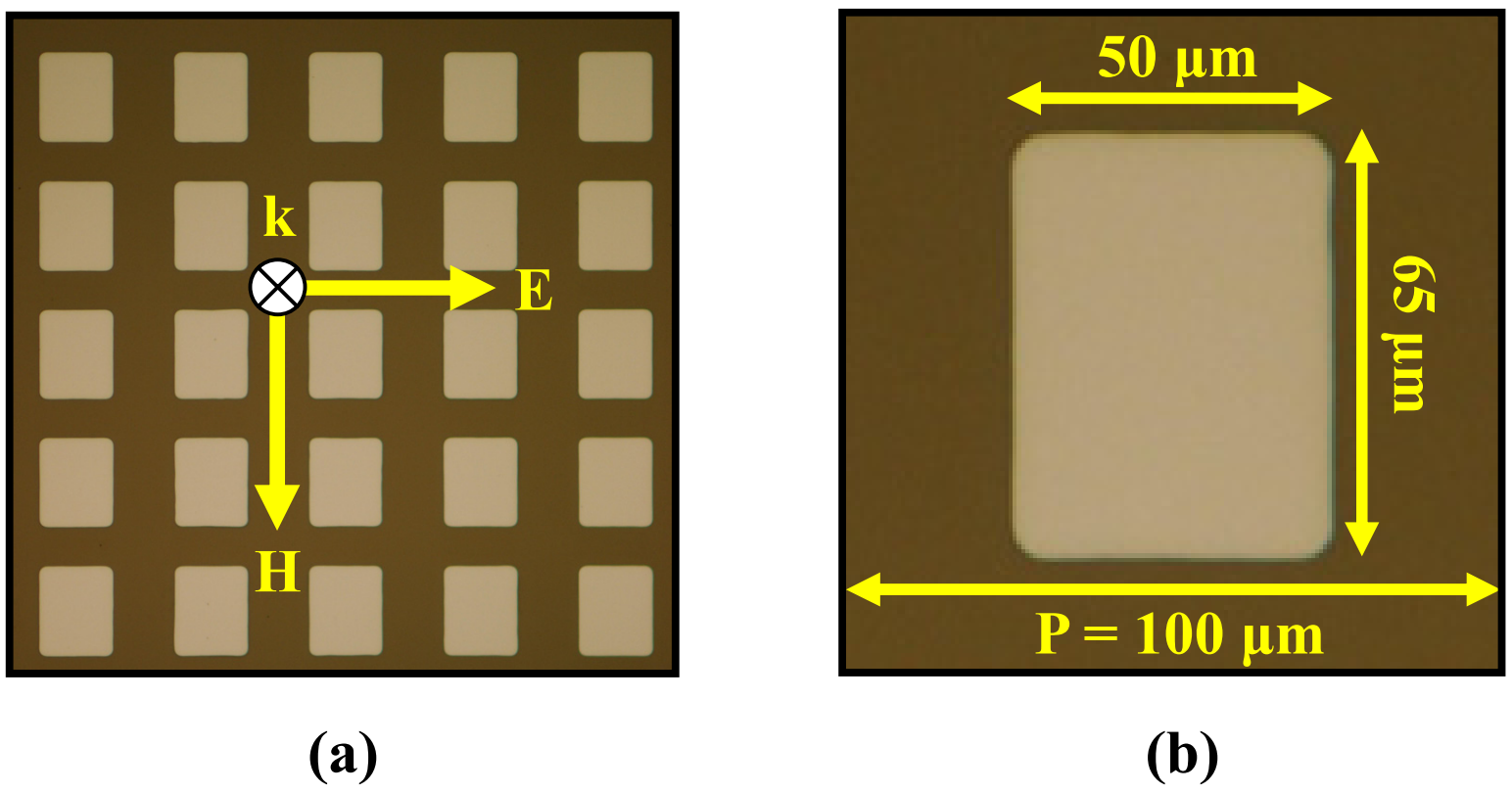

**FIG. 1.**
**Tian *et al.***

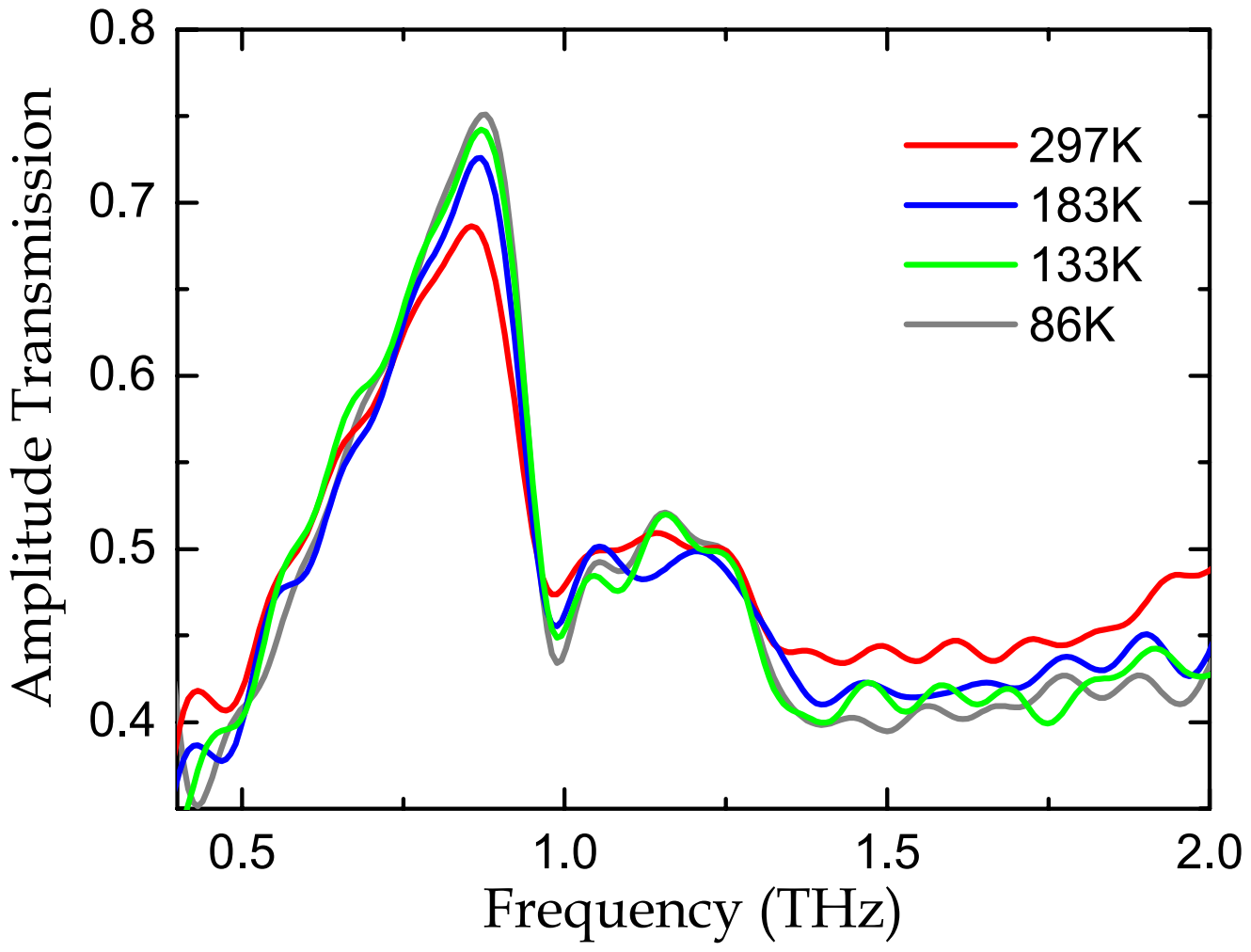

**FIG. 2.**
**Tian *et al.***

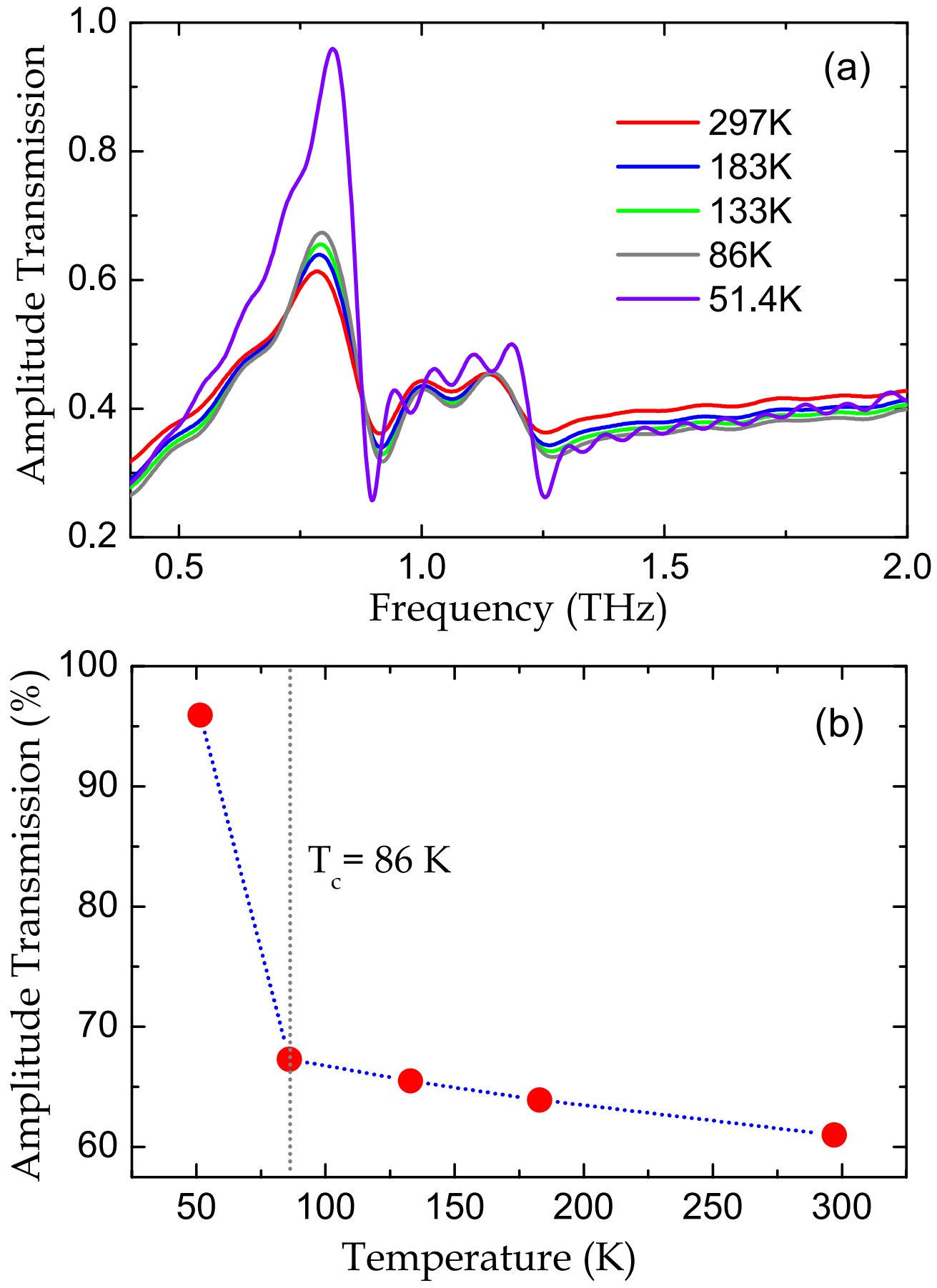

**FIG. 3.**
**Tian *et al.***